\documentclass[%
 aip,
 sd,%
 amsmath,amssymb,
 reprint,%
]{revtex4-1}

\usepackage{graphicx}
\usepackage{dcolumn}
\usepackage{bm}
\usepackage{xcolor} 

\begin{document}

\preprint{AIP/123-QED}

\title[Letter title]{Critical angle for interfacial phonon scattering:\\ results from \textit{ab initio} lattice dynamics calculations}

\author{A. Alkurdi}
 \altaffiliation[Also at ]{ Department of Physics, Al-Baath University, Homs, Syria}
\email{ali.alkurdi@univ-lyon1.fr}
\author{S. Pailh\`es}%
\affiliation{ Institut Lumi\`{e}re Mati\`{e}re, UMR5306 Universit\'{e} Claude Bernard Lyon 1-CNRS, Universit\'{e} de Lyon 69622 Villeurbanne Cedex, France
}%

\author{S. Merabia}
 \email{samy.merabia@univ-lyon1.fr}

\date{\today}

\begin{abstract}
Thermal boundary resistance is a critical quantity that controls heat transfer at the nanoscale, and which is primarily related to interfacial phonon scattering. Here, we combine lattice dynamics calculations and inputs from first principles \textit{ab initio} simulations to predict phonon transmission at Si/Ge interface as a function of both phonon frequency and phonon wavevector. This technique allows us to determine the overall thermal transmission coefficient as a function of the phonon scattering direction and frequency. Our results show that, the thermal energy transmission is highly anisotropic while thermal energy reflection is almost isotropic. In addition, we found the existence of a global critical angle of transmission beyond which almost no thermal energy is transmitted. This critical angle around $50^{\circ}$. is found to be almost independent on the interaction range between Si and Ge, the interfacial bonding strength, and the temperature above $30$ K. We interpret these results by carrying out a spectral and angular analysis of phonon transmission coefficient and differential thermal boundary conductance.
\end{abstract}

\maketitle

\section{Introduction}
Thermal transport at the nanoscale is at the heart of modern microelectronics. With the continuous reduction of the size of the microprocessors, 
there is a strong need to understand heat flow across interfaces from a fundamental standpoint. Indeed, in the modern microelectronic era, transistors reach nanometer length scales, and their relevant dimensions become comparable with the Kapitza length $l=\lambda/G$ where $\lambda$ is the thermal conductivity and $G$ the thermal boundary conductance (TBC) characterizing the interface. Typical values for the Kapitza length are between $1$ and $100$ nm, showing the importance of thermal interface resistance for nanostructured materials~\cite{Pop2010}.

In conventional metal/dielectrics interfaces, heat transport is mediated by both phonon-phonon and electron-phonon scattering~\cite{Lombard2015}. For interfaces between semiconductors, only phonon-phonon scattering is relevant, but even this simpler situation is far from being understood. In particular, the thermal energy transmission coefficient has not been measured experimentally and thermal process at interfaces remain poorly understood. Notable exceptions may be found where phonon reflection and internal reflection have been evidenced experimentally~\cite{Northtrop1984,Höss1990}, but these manifestations concern sub-Thz frequencies, a situation where continnum acoustics and thus Snell-Descartes laws should be at work. In contrast to phonon scattering, reflection and transmission coefficients for {\em photon} interfacial scattering are well described by Fresnel coefficients~\cite{Hua2015}. Even the involved problem of the intensity scattered by a rough interface has been adressed theoretically by Maradudin and coworkers~\cite{Simonsen2010}. Such a milestone has not been reached for phonon scattering and for thermal transport at crystalline interfaces in general. The reasons are manifold: i) crystals are anisotropic; ii) phonon dispersion is not as simple as electromagnetic dispersion; iii) phonon transmission coefficient should depend on the interfacial bonding strength between the two solids.   

In order to rationalize early experiments on thermal boundary resistance at low and moderate temperatures, phenomenological models have been 
proposed, including the popular acoustic mismatch model (AMM)~\cite{Little1959,Chen1999} and the diffuse mismatch model (DMM)~\cite{Swartz1989}. However, these two models generally fail in describing accurately interfacial thermal transport as they do not include any information relevant to the interface, and usually consider phonon dispersion in a simplified way. In the quest for more accurate models, several computational methods have been proposed during the last decade, including molecular dynamics~\cite{Landry2009}, Green's function~\cite{Zhang2007} and lattice dynamics~\cite{Young1989,Zhao2005}. 

Molecular dynamics is a flexible tool to describe bulk phonon properties and interfacial thermal transport, and its accuracy would be considerably improved by the use of interatomic potentials deriving from ab-initio calculations.
Green's function calculations may challenge these limitations, but they have not been devised to compute the frequency and wavevector dependence of the transmission coefficient, at least for an interface between two 3D materials~\cite{Ong2015}.

In this Letter, we combine lattice dynamics calculations with {\it ab initio} interatomic force constants in order to predict phonon transmission as 
a function of {\em both} phonon frequency and scattered angle, for a Si/Ge interface. Our results show that phonon transmission is highly anisotropic, while phonon reflection is almost isotropic. We demonstrate that almost all the thermal energy is transmitted at the interface for scattered angles smaller than a critical value which is found to be typically $50\,^\circ$ for the system considered here. Through the consideration of different numbers of interacting atoms at the interface, we systematically studied the effect of the interfacial interaction range and the interfacial bonding strength, and conclude that these factors mildly affect the angular distribution of phonon scattering. In contrast,  temperature is found to have more influence, as a result of the phonon occupation statistics.   

\section{Theoretical Model}
The model we developed is inspired by the lattice dynamics model for Si/Ge interface, as described in \cite{Zhao2005}. We have extended this latter model basically to employ interatomic force constants inputs from {\it ab initio} calculations. In this perspective, it was necessary to modify the equations of motion detailed in~\cite{Zhao2005} so as to account for the interaction with $8$\textsuperscript{th} neighbouring atoms, while the original work by 
Zhao and Freund considers an empirical Stillinger-Weber potential which is short range and simpler to tackle, but which lacks the accuracy to describe phonon bulk dispersion curves especially for Ge. 
The interaction in each medium is supposed to be determined by interatomic force constants $\mathbf{\Phi}(\mathbf{r})$ derived from {\it ab initio} calculations~\cite{Aouissi2006} that have been calculated employing density functional 
pertubation theory, a pseudopotential approach and the local density approximation (LDA) for the exchange-correlation potential. As shown in the 
supplementary material, the phonon spectra are in good agreement with experimental data for Si and Ge. Note that the phonons dispersions curves obtained with the force constants described above doesn’t not show negative phonon branch. 

\begin{figure}[htbp]
\includegraphics[scale=0.25]{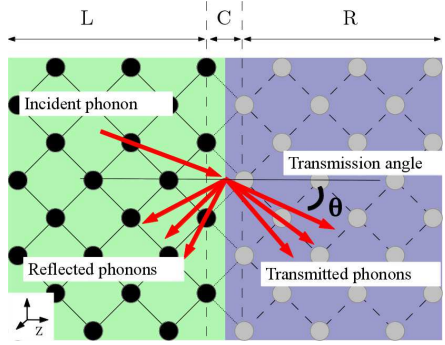}
\caption{\label{fig:system} Schematic of the perfect interface between Si/Ge. Coming from the left lead, the incident phonon ($\mathbf{k}$) strikes the interface in the central region and is decomposed into transmitted phonons in the right lead ($\mathbf{k}^R$) and reflected phonons in the left lead ($\mathbf{k}^L$).}
\end{figure}

 Under the harmonic approximation, the equation of motion for an atom $i$ of mass $m_i$ displaced by $\mathbf{u}(\mathbf{r}_{i})$ can be read as~:~$ m_i\mathbf{\stackrel{..}{u}}(\mathbf{r}_i)=-\sum_{\acute{i}}{ \mathbf{\Phi}(\mathbf{r}_{i\acute{i}})\mathbf{u}(\mathbf{r}_{\acute{i}})}$
where the sum is over all atoms including atom $i$ itself and $\mathbf{\Phi}$ is the interatomic force constant matrix limited here to harmonic terms. 
The equation of motion can be rewritten as~:~$\mathbf{D}(\mathbf{k})\mathbf{e}(\mathbf{k},\nu)=\omega^2(\mathbf{k},\nu)\mathbf{e}(\mathbf{k},\nu)
 \label{eq:dynamicsInBulck}$ where $\bf{k}$ is the wavevector, $\nu$ denotes the branch mode and $\mathbf{D}$ is the dynamical matrix here having dimensions $6\times 6$.
The interface is represented in Fig.~\ref{fig:system}, where we divide the system into three regions: the left lead where the incident phonons come from, the central region corresponding to the location of the interface and the right lead where the transmitted phonons come out, denoted by L, C and R, respectively. The interface is oriented to be perpendicular to the $z$-axis corresponding to the [001] crystalline direction. 
The dynamical matrix can thus be decomposed into three or more region-depending dynamical matrices as~\cite{Zhao2005}:
$\mathbf{D} = \Gamma^{-1}\mathbf{D}^L +\mathbf{D}^C_S + \Gamma\mathbf{D}^R$ where $\Gamma = \exp{(ik_za/2)}$ and $z$ denotes the direction normal to the interface. We assume elastic scattering at the interface, an incident phonon can only excite phonons having the same frequency $\omega(\mathbf{k}^L)=\omega(\mathbf{k}^R)=\omega(\mathbf{k})$ and the components of the wave vector parallel to the interface are conserved~\cite{Pettersson1990}  $k^L_x=k^R_x=k_x \hspace{0.5cm} \textmd{and} \hspace{0.5cm}  k^L_y=k^R_y=k_y$.
 These equations have complex displacement vectors as solution, $\mathbf{u}(\mathbf{r}_{ij})$ of the $i$\textsuperscript{th} atom in the $j$\textsuperscript{th} primitive unit cell in the left lead which can be expressed as a superposition of the incident mode and the excited modes in the left lead~:
\begin{equation}
 \mathbf{u}(\mathbf{r}_{ij})=\mathbf{A}_i(\mathbf{r}_{i},\mathbf{k}, \nu) \exp\left(i\mathbf{k}.\mathbf{r}_{j}\right) +\sum^{n_r}_{\mathbf{k}_r, \nu}{\mathbf{A}_r(\mathbf{r}_{i},\mathbf{k}_r, \nu) \exp \left(i\mathbf{k}_r.\mathbf{r}_{j}\right)}  \label{eq:solutionL}
\end{equation}
and similarly, the displacement vector in the outcoming medium (right lead) is 
\begin{equation}
 \mathbf{u}(\mathbf{r}_{ij})=\sum^{n_t}_{\mathbf{k}_t, \nu}{\mathbf{A}_t(\mathbf{r}_{i},\mathbf{k}_t, \nu) \exp \left(i\mathbf{k}_t.\mathbf{r}_{j}\right)}  \label{eq:solutionR}
\end{equation}
where $\mathbf{A}_r$, $\mathbf{A}_t$ are the amplitudes of the reflected and transmitted waves, $n_r$ and $n_t$ are the numbers of allowed reflected and transmitted modes, respectively (here $n_r;n_t \ge 6$ for the diamond like systems considered).
By solving this set of equations at the interface, we obtain the amplitude of each excited mode leading to the knowledge of the individual mode-dependent energy transmission coefficient which is defined as the fraction of phonon energy transmitted across the interface~: 
\begin{equation}
  \mathcal{T}(\mathbf{k}_t,\nu)= \frac{\rho_t}{\rho_i}\frac{v^{z}_{g,t}.\vert A_{t}\vert^2}{v^{z}_{g,i}.\vert A_{i}\vert^2} \label{eq:tdef}
\end{equation}
where $\rho_i$ and $\rho_t$ are the mass density in the incidence and transmission medium, respectively. $v^{z}_{g,t}$ are the group velocities of the transmitted phonons projected along the direction perpendicular to the interface. Similarly, the individual reflection coefficient is~: 
\begin{equation}
  \mathcal{R}(\mathbf{k}_r,\nu)= \frac{v^{z}_{g,r}.\vert A_{r}\vert^2}{v^{z}_{g,i}.\vert A_{i}\vert^2} \label{eq:rdef}
\end{equation}
The mode-dependent thermal conductance has the expression~:
\begin{equation}
G(\mathbf{k}_t,\nu) = \frac{1}{V}  v^{z}_{g,t} \hbar \omega \frac{\partial f(\omega,T)}{\partial T} \mathcal{T}(\mathbf{k}_t,\nu)
\end{equation}
where $V$ is the volume of the system and $f$ denotes the phonon occupation density and the sum is over the transmitted modes.
In this work, the {\em ab-initio} interatomic force constants are taken from Ref.~\cite{Aouissi2006}. Unless specified, we consider that the interatomic force constants at the interface are equal to the arithmetic mean of Si and Ge force constants, which can be summarized as~: $\Phi_{Si-Ge}=
\frac{1}{2} (\Phi_{Si-Si}+\Phi_{Ge-Ge})$ with obvious notations. We note that our approach neglects anharmonic effects as a result of the elastic interfacial scattering assumption, but as noticed in~\cite{Tian2012,Landry2009} anharmonic effects become important for Si/Ge at temperatures higher than $500$ K. 

To critically assess the method, we have compared the total thermal conductance $G$ calculated using the transmission coefficient and found a value $G = 170$ (MW/m\textsuperscript{2}/K) at high temperatures. This value is in good agreement with the conductance reported in 
a DFT-Green's function 
calculation for the same interface~\cite{Tian2012}, thus validating our methodology. Note also, that the resulting value of the TBC is smaller by a factor $2$ than the value predicted by the empirical Stillinger-Weber potential~\cite{Landry2009,Zhao2005}, showing the necessity to employ ab-initio interatomic force constants.   

\section{Results and Discussion}
\begin{figure}[htbp]
\centering
\includegraphics[scale=0.25]{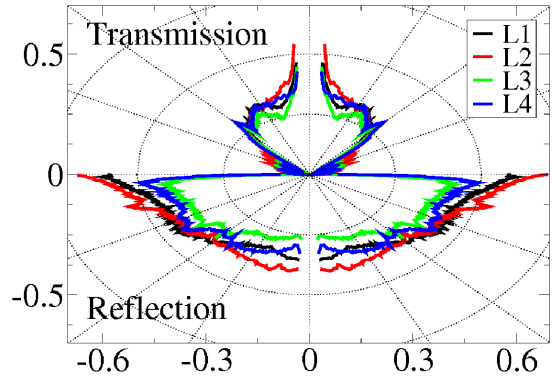}
\caption{\label{fig:T_polar_SiGe}(Color online) (a)~Phonon energy transmission at Si/Ge in polar coordinates for different interaction ranges. 
L1 refers to the interaction limited to only the nearest unit cell while L4 refers to the interaction range up to the 4th unit cell.}
\end{figure}

We present now the results obtained by performing the LD calculations on Si/Ge interface described in the previous section. The interatomic interaction was limited to the $4$\textsuperscript{th} unit cell~(corresponding to $8$\textsuperscript{th} nearest neighbours for diamond-like crystals). Fig.~\ref{fig:T_polar_SiGe} shows the transmission and reflection of all phonon modes in polar coordinates, and for different interaction ranges that goes from the first unit cell (L1) up to the $4$\textsuperscript{th} unit cell (L4). It is important to mention that we have performed an intensive statistical treatment, and considered $3$ millions excited incident wavevectors. The scattering angle is defined using group velocities $\cos(\theta)=v^z_g/\mathbf{v}_g$. Overall, the transmission coefficient displays a maximum value close to $1/2$ in the normal transmission direction corresponding to $\theta=0\,^\circ$ and it decreases when we go to the grazing direction, $\theta=90\,^\circ$. From this analysis, we can see that the phonon transmission is anisotropic with relative low values lying in the interval of transmission angle between  
$\theta=50\,^\circ$ and $\theta=90\,^\circ$. This transmission angle $\theta^c_{a}=50\,^\circ$ beyond which the transmission is small can be seen as a global critical angle for phonon scattering. Thermal phonon reflection is to be almost isotropic with maximum value about $0.7$ in the direction of $90\,^\circ$ of transmission angle. The transmission coefficient is found to be almost insensitive to the interaction range, but the reflection coefficient displays little variations when the interaction goes up to the $4$\textsuperscript{th} unit cell.

\begin{figure}[h]
\includegraphics[scale=0.6]{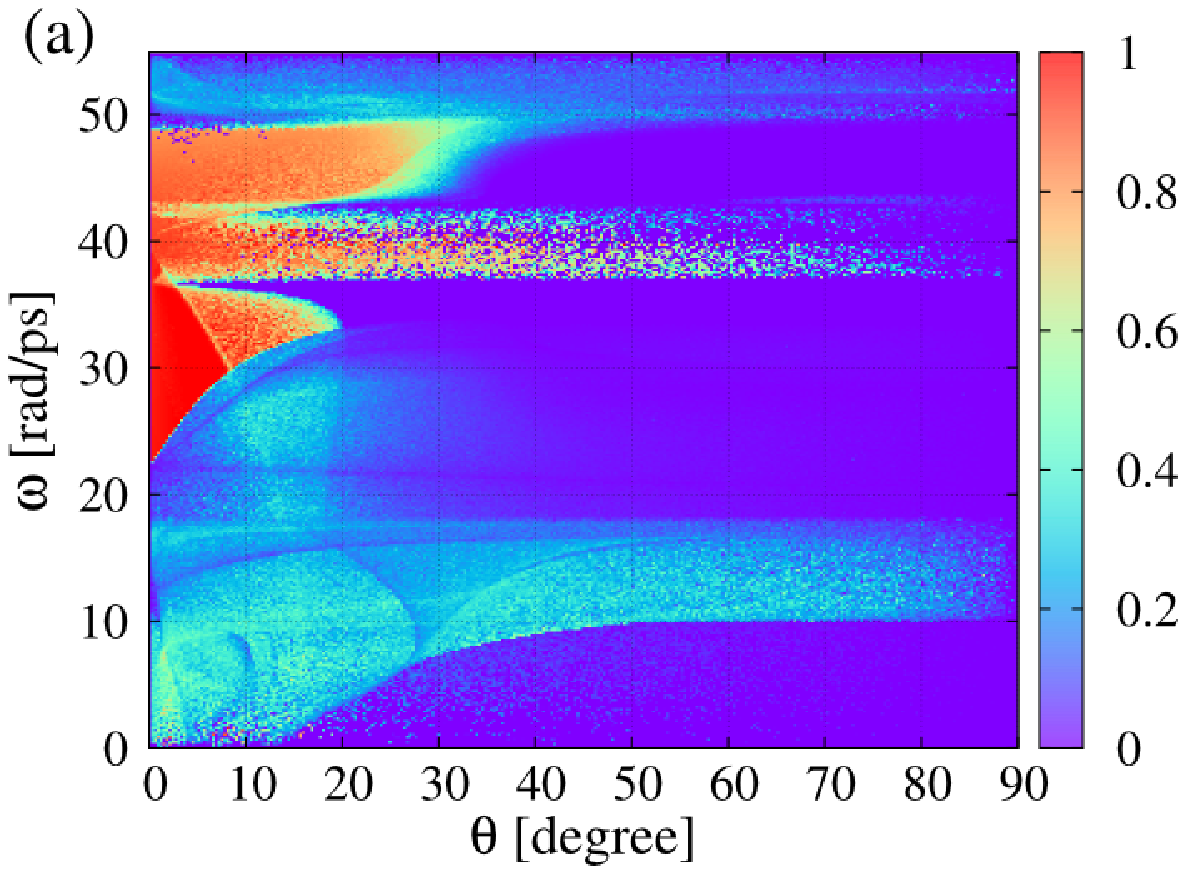}
\includegraphics[scale=0.6]{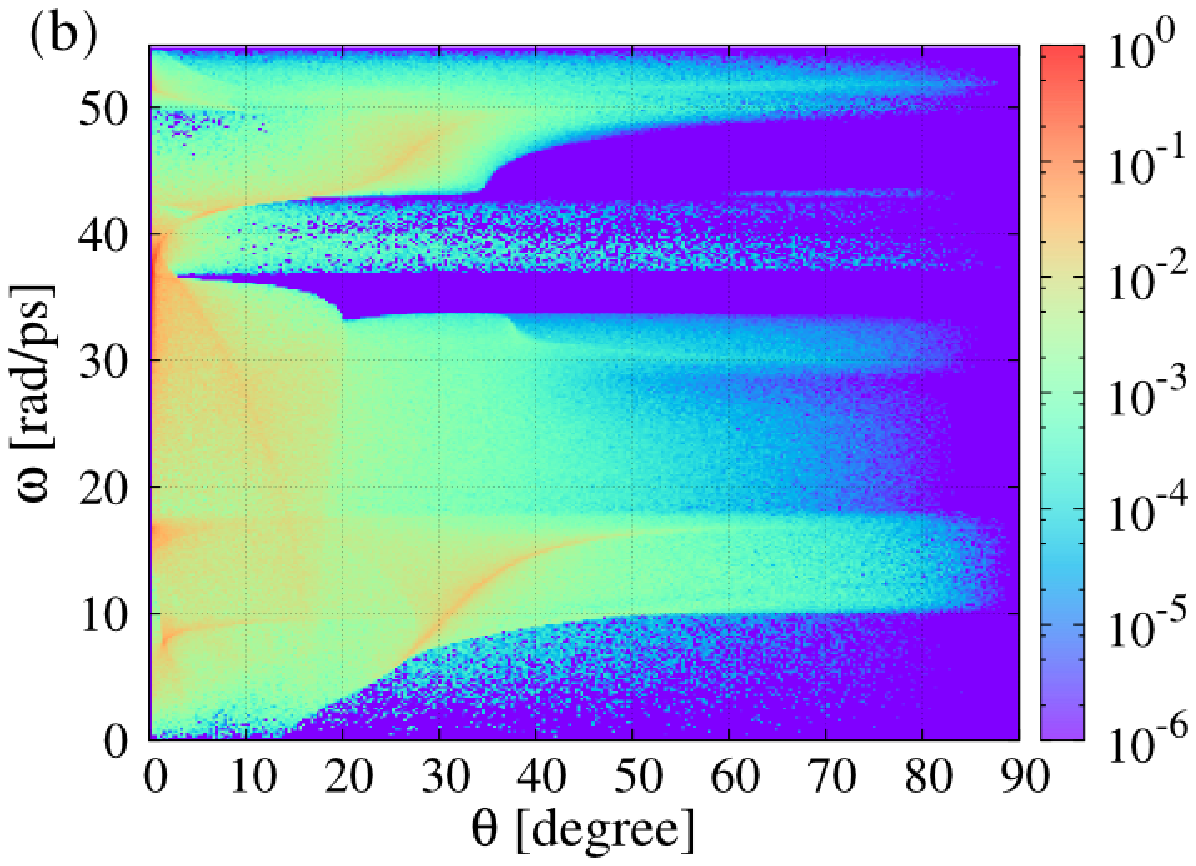}
\caption{\label{fig:Si_Ge_w_cos}(Color online) (a)~:~Colormap of the phonon energy transmission at Si/Ge interface as a function of both the angular frequency and the transmission angle. 
(b)~:~Same for the thermal boundary conductance (in units of MW ps/m\textsuperscript{2}K rad). Note the log scale used in the color map.
The interaction range was taken up to the $2$\textsuperscript{nd} unit cell here.}
\end{figure}

We have performed a spectral and angular analysis to quantify the contribution of each phonon mode in a given scattering direction, 
as depicted in Fig.~\ref{fig:Si_Ge_w_cos}. Again, this figure presents the result of $3$ million excited modes. We can split the phonon frequency range, which extends up to the Ge cut-off frequency, in three domains.
For low frequency phonon modes $\omega<20$ (rad/ps) the transmission probability is small and we can clearly see the existence of a critical angle around $12^\circ$ for zero frequency modes. Intermediate frequency modes, $20<\omega<45$ (rad/ps), corresponding to longitudinal acoustic modes of Germanium have a large probability to be transmitted for small scattered angles only, $\theta<20^\circ$. The corresponding contribution of heat flux is high, as can be inferred from Fig.~\ref{fig:Si_Ge_w_cos} b). Lastly, Germanium optical modes having frequencies  $\omega>45$(rad/ps) are characterized by a large transmission coefficients for angles $\theta<30^\circ$, but the amount of heat they carry is small. Hence, these modes 
only contribute to the heat capacity without transmitting significant heat flux.
   
\begin{figure}[h]
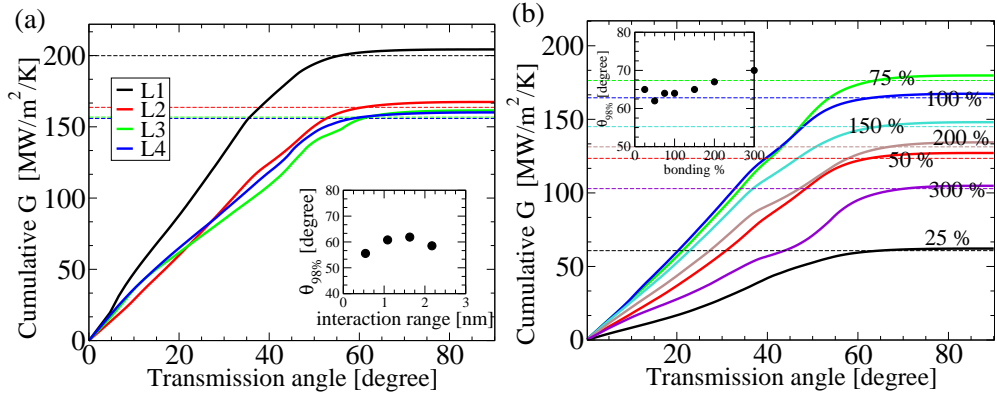

\includegraphics[scale=0.25]{GCuml_theta_SiGe.eps}
\includegraphics[scale=0.25]{GCuml_theta_SiGe_Adj.eps}
\caption{\label{fig:GCuml_theta_SiGe}(Color online) The cumulative TBC at room temperature for Si/Ge interface as a function of transmission angle for different interaction ranges (a) and different interfacial bonding strengths~(b) taken as a percentage of the mean value of the bonding in the bulk of both media (Si and Ge). The insets show the critical angle $\theta_{98\%}$ corresponding to 98\% of the total transmitted energy.}
\end{figure}

\begin{figure}[h]
\includegraphics[scale=0.25]{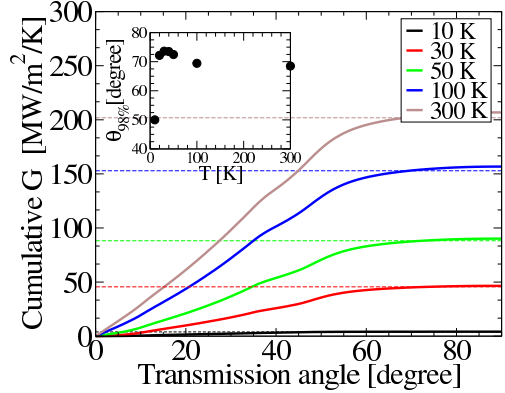}
\caption{\label{fig:G_theta_SiGe_T}(Color online) The cumulative TBC of Si/Ge interface as a function of transmission angle for different temperatures. The inset shows the critical angle $\theta_{98\%}$ corresponding to 98 \% of the total transmitted energy as a function of temperature.}
\end{figure}

In Fig.~\ref{fig:GCuml_theta_SiGe}, we plotted the cumulative TBC as a function of transmission angle for different interatomic interaction ranges (a) and for different interfacial interaction strengths (b). In both figures, we see again that the significant contribution to the energy transmission across the interface comes from the angular directions corresponding to transmission angles smaller than $60\,^\circ$. The TBC exhibits a significant drop in magnitude when the interfacial interaction range goes beyond the first unit cell (L1). This behaviour may be surprising and contrary to the intuition, nevertheless, it demonstrates that increasing the range of the interaction does not necessarily imply enhanced phonon transmission. The conductance converges when the interfacial interaction range spans more than three unit cells.
If we now examine the role of the interaction strength, taken as a percentage of the mean value of the bonding in the bulk of both media (Si and Ge), Fig.~\ref{fig:GCuml_theta_SiGe}(b) shows that the TBC increases with the interfacial interaction strength up to a value where we have a maximum and then the TBC starts to decrease for larger interfacial bonding strengths. 
 We are interested to know up to which direction 98\% of thermal energy is transmitted to the outcoming medium (Ge) and how the corresponding critical angle $\theta_{98\%}$ changes with respect to the interfacial bonding strengths and the interatomic interaction range as well. To this end, we plotted the limiting angle in the insets of Fig.~\ref{fig:GCuml_theta_SiGe}~(a) and (b). We found that the variations of this limiting angle with respect to the interatomic interaction range are not that large. Moreover, the limiting angle is found to be not so sensitive to the interfacial bonding strength, but when the relative bonding strength is larger than 100~\% , the corresponding limiting angle slightly levels off.
Finally, we illustrate in Fig.~\ref{fig:G_theta_SiGe_T} the temperature dependence of the cumulative thermal boundary conductance as a function of the transmission angle. It is obvious that the TBC increases with respect to temperature and saturates at high temperatures. The inset shows the variations of the limiting angle with respect to the temperature. The limiting angle clearly increases with temperature and saturates for a temperature higher than $30$ K.
The increase is related to the phonon statistics and the progressive population of high frequency phonon modes which may transfer energy 
at large scattering angles , as can be inferred from fig.~\ref{fig:Si_Ge_w_cos}.


%
%
\section{Conclusion}
In summary, we have used LD calculations combined with {\it ab initio} interatomic force constants to investigate thermal transport at silicon/germanium interface. These calculations have allowed us to quantify the transmission and reflection coefficients as a function of 
scattered angle, enlighting the important contribution of phonons transmitted along the normal direction. A spectral analysis has been carried out to elaborate a complete view on the contribution of each phonon mode in a given scattering direction. We have also demonstrated the existence of a limiting scattering angle beyond which thermal transmission is negligible. This limiting angle has been shown to depend mildly on the interatomic interaction range and the interfacial bonding strength, but increases with temperature. 
Further improvement of the methodology, applied here to the case of the Si/Ge interface, will need to address more precisely the force constants at the interface by means of ab initio calculation in large supercells, especially for interfaces between materials that do not have the same bulk crystalline structure. This work may be considered as a first step towards the development of multi-scale theoretical analysis of phonon transport in real devices with nanostructured interfaces~\cite{Merabia2014}. 


\section{Supplementary material}
See supplementary material for the phonon spectra of Silicon and Germanium calculated with the ab initio force constants.

\section{Acknowledgements}
The authors acknowledge fruitful discussions with P.-O. Chapuis, C. Adessi, R. Viennois and K. Termentzidis. Partial financial support from the 
ANR Mascoth is also acknowledged.

\end{document}